# Robust and efficient quantum optimal control of spin probes in a complex (biological) environment. Towards sensing of fast temperature fluctuations.

Philipp Konzelmann [*,a], Torsten Rendler [*,+,a], Ville Bergholm [b], Andrea Zappe [a], Veronika Pfannenstill [a,c], Marwa Garsi [a], Florestan Ziem [a], Matthias Niethammer [a], Matthias Widmann [a], Sang-Yun Lee [d], Philipp Neumann [+,a], Jörg Wrachtrup [a]

[a] *3. Physikalisches Institut, Center for Applied Quantum Technologies and IQST, Pfaffenwaldring 57, 70569 Stuttgart, Germany*

[b] *Dept. of Chemistry, Technical University of Munich (TUM), 85747 Garching, Germany*

[c] *Institute of Applied Physics, TU Wien, Wiedner Hauptstrasse 8-10, Vienna 1040, Austria*

[d] *Center for Quantum Information, Korea Institute of Science and Technology, Seoul 02792, South Korea*

[+] *Inquiries should be sent to one of the following e-mail addresses:*

*Torsten Rendler: t.rendler@physik.uni-stuttgart.de*

*Philipp Neumann: philipp@nvision-imaging.com*

** These authors contributed equally*

ABSTRACT

We present an optimized scheme for nanoscale measurements of temperature in a complex environment using the nitrogen-vacancy center in nanodiamonds. To this end we combine a Ramsey measurement utilized to temperature determination with advanced optimal control theory. We test our new design on single nitrogen-vacancy centers in bulk diamond and fixed nanodiamonds, achieving better readout signal than with common soft or hard microwave control pulses. We demonstrate temperature readout using rotating nanodiamonds in an agarose matrix. Our method opens the way to measure temperature fluctuations in complex biological environment. The used principle however, is universal and not restricted to temperature sensing.

## INTRODUCTION

Spin quantum probes have the potential to measure a wealth of parameters with unprecedented accuracy and spatial resolution. One representative of such quantum sensors, is the negatively charged nitrogen-vacancy center (NV) in diamond. Several applications as a magnetic[1] and electric field,[2] pressure[3] and temperature[4–7] probe have been demonstrated. In addition, one can also extract chemical information about samples by nuclear magnetic resonance (NMR) with ~Hz spectral and nanometer-scale spatial resolution.[8,9] These techniques can be modified to measure the temporal dynamics of the quantity under study using NV. [10,11]

The sensor capability of NV can be preserved in nanodiamonds (NDs), even in a biological environment,[12] where spin coherence and relaxation times are greatly reduced. As the NV is a spin one system (S=1), the response to external magnetic fields is anisotropic: tumbling of the host ND will lead to variations in the electron spin resonance (ESR) transitions strength and frequency. Both depend on the orientation of the NV axis (the defect symmetry axis) with respect to the external static magnetic and microwave field. Therefore, proper alignment of the NV spin system is usually mandatory to perform precise quantum measurements.

To counteract variations in the NV spin transitions frequency and its driving strength, one can utilize optimal control theory, which has been used in a manifold of variations for NV. [13–17] Thereby one can numerically optimize a pulse to achieve the desired quantum state.

To perform the optimization we use the quantum optimization package DYNAMO, [18] which uses the gradient ascent pulse engineering (GRAPE) algorithm together with Broyden-Fletcher–Goldfarb-Shanno (BFGS) minimization. DYNAMO can also optimize open quantum system dynamics. By introducing cooperativity between pulses of a given sequence *via* the quantum state filter method by Braun and Glaser,[19] possible errors in phase adjustment can be compensated by the following pulses. Overall this reduces the demand for individual pulses, while still achieving the overarching goal of the entire sequence utilizing less resources like time or energy.[19] To demonstrate the power of the used optimization scheme, we modified the dynamical decoupling sequence D-Ramsey, optimized for temperature sensing with NV,[4] to a new sequence, which allows us to monitor local temperature even with tumbling nanodiamonds.

The original D-Ramsey involves the allowed ESR transitions $|0\rangle \leftrightarrow |+1\rangle$ and $|0\rangle \leftrightarrow |-1\rangle$ of the triplet ground state of the NV center (see Figure 1A).

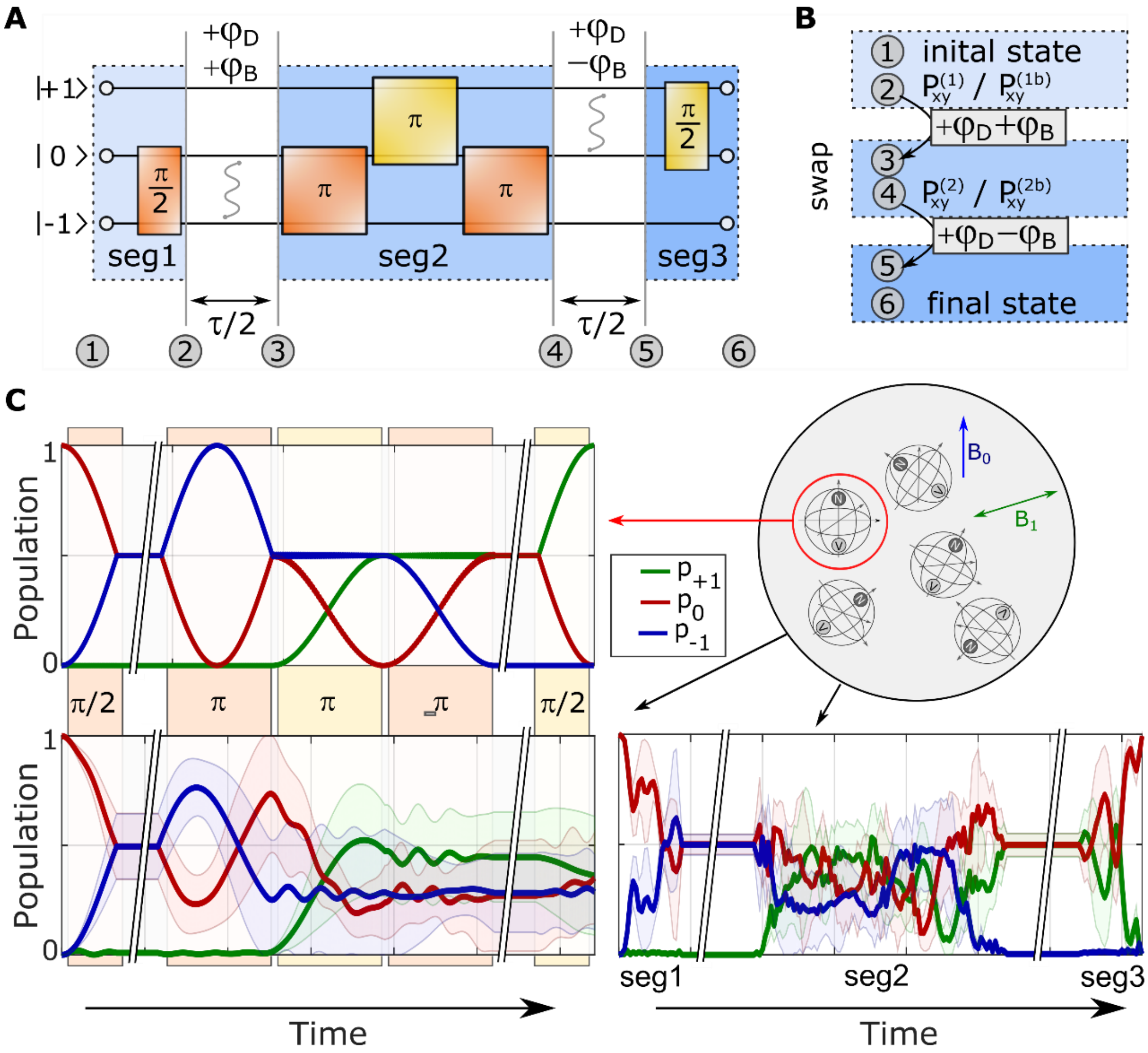

**Figure 1. Working principle of D-Ramsey. (A):** Translation of D-Ramsey into a cooperative design. The sequence is sliced into 3 segments (seg1, seg2 and seg3), and each segment is separated equally by $\tau/2$.$\varphi_D$ and $\varphi_B$ denote quantum phase accumulation due to changes in $D$ or the external magnetic field. **(B):** Before and after seg2 projectors are applied to introduce a cooperative design in the pulse compilation via optimal control. **(C):** Example of state evolution within the sequence for different drift Hamiltonians and different scenarios. The upper and lower left plots show the state evolution under square pulses with fixed system parameters and a various amplitude and detuned ensemble, respectively. The lower right shows the state evolution under a Coop-D-Ramsey sequence optimized for this ensemble. Thereby, for the lower plots, the solid lines present the average state of the ensemble. The corresponding colored areas are the spreading of state evolutions in terms of one sigma. $B_0$ and $B_1$ are the applied external magnetic field vector and microwave field, respectively.

These two ESR lines can shift due to changes in many internal and external parameters. For example, if the zero field splitting $D$ of the NV changes by local temperature changes,[20] they shift in the same direction. A change of the strength of an applied axial magnetic field $B_0$, shift the ESR lines in opposite directions. The D-Ramsey is designed in a way, that only the center of gravity (COG) of both transitions is measured: hereby the sequence acts as a Ramsey accumulating the phase $\varphi_D = \delta D \cdot \tau$ during the free evolution interval $\tau$, if the COG is detuned by $\delta D$. Slow changes in $B_0$, introducing a possible phase shift $\varphi_B$, are canceled out similar to a Hahn-Echo sequence. One can separate the D-Ramsey into three pulse segments with individual roles (see Figure 1A and B). 'seg1' initializes quantum sensing, by creating a superposition between $|0\rangle$ and $|-1\rangle$. 'seg2' prepares the system to be only sensitive to COG shifts of $|-1\rangle$ and $|+1\rangle$, which is accomplished by swapping the populations of $|-1\rangle$ and $|+1\rangle$.'seg3' converts phase to state population enabling readout.

To understand how we translate the D-Ramsey scheme into a cooperative pulse design, we quickly recall the basic idea of optimal control theory: the dynamics of a quantum system with density matrix $\rho$ can be described by the von Neumann equation:

$$\dot{\rho}(t) = -i \left[ \underbrace{H_{\text{drift}} + \textstyle\sum_k u_k(t) H_k}_{H}, \rho(t) \right], \tag{1}$$

with $H_{\text{drift}}$ being the free evolution or drift Hamiltonian, while $H_k$ are the control Hamiltonians that describe the interaction of the system with for example an external microwave field. For magnetic fields, $H_k$ is composed of the spin operators $S_x$,$S_y$ and $S_z$. The amplitude $u_k(t)$ at time $t$ of this interaction is typically called control field. The task is to find all $u_k(t)$, such that the system evolves into a predefined final state after time $T$, represented by the density matrix $\lambda$. Therefore, one defines a quantity $\Phi_0$ that can for example be the overlap of the desired final state $\lambda$ with the actual state $\rho$ evolved under the influence of the control fields: $\Phi_0 = \langle \lambda | \rho(T) \rangle$. By maximizing $\Phi_0$, one optimizes the amplitude $u_k(t)$ to the designed functionality. Hereby one of the strengths of optimal control lies in the possibility to find a particular set of $u_k(t)$ that allows robustness against variations of system parameters (*e.g.* a broad range of $H_{\text{drift}}$ and $H_k$). [21]

To introduce cooperativity using the quantum state filter method, we describe the evolution of the system in the Liouville space. The evolution of a spin system with density matrix $\rho_s$ that is coupled to an environment can be described as:

$$\rho_s(t_2) = \exp(\mathcal{L}(t_1, t_2)\Delta t)\, \rho_s(t_1). \tag{2}$$

Hereby $\mathcal{L} = -\mathrm{i} \cdot \mathcal{H} + \mathcal{D}$ is the Liouville superoperator, being the sum of the time independent system Hamiltonian $\mathcal{H}$ and the dissipator $\mathcal{D}$ in the time interval $\Delta t = t_2 - t_1$. The exponential term in eq (2) is typically called a dynamic map $P$. In a different perspective, $P$ can also be understood as a projector. By replacing the dynamic map of the system, evolving with $\mathcal{H}$ between each segment with a suitable projector, the full sequence can be compiled at once invoking cooperativity. Thereby $P$ projects on a subspace in the Liouville space of the density matrices, which in our case is any equal superposition of states $|0\rangle$ and $|-1\rangle$ after 'seg1' and $|0\rangle$ and $+|1\rangle$ after 'seg2'. If the state before applying the projector does not belong to this subspace, some purity is irreversibly lost and the fidelity of the final state reduces. This causes DYNAMO to optimize the amplitudes $u_k(t)$ such that the desired subspace is reached before the projector is applied. After optimization, we can simply cut the overall sequence between the segments and add arbitrary but equally long free evolution interval in between.

As shown in the following section, the introduced strategy allows us to measure the local temperature of a ND, despite its rotation, which would result in incoherent control using the standard sequence. Note, that in principle, one can apply this recipe to any measurement scheme (in particular multi-phase dynamic decoupling sequences).

## RESULTS AND DISCUSSION

As we stated in the introduction part, optimal control theory can be utilized to introduce a certain robustness against parameter variations. For demonstration we use the sequence compiled for later study, which we referee to as bulk case (see also Supporting Information (SI)). Hereby we compare the Coop-D-Ramsey (cooperatively numerically optimized D-Ramsey) to a typical D-Ramsey consisting of conventional π and π/2 pulses. Figure 1C shows the population of all three states for different $H_{\mathrm{drift}}$ and $H_k$, for three different situations. The definition of $H_{\mathrm{drift}}$ and $H_k$ can be found in eqs (S2b,c,d). The upper left plot pictures the population of all three states for a well-set ideal parameter choice in case of a D-Ramsey. The sequence works as desired. The lower plots represent the average over a sample of 63 different sets (for parameters see bulk study and Methods) of $H_{\text{drift}}$ and $H_k$ for a D-Ramsey and Coop-D-Ramsey sequence. As one can see, for both, D-Ramsey and Coop-D-Ramsey, the population has a high divergence within the individual segments. However, the Coop-R-Ramsey converges at the end of each segment to the desired superposition state. To calculate the optimal set of $u_k(t)$ for the Coop-D-Ramsey, we used the projectors, which are described in eqs (S4a,b) in the SI. We labeled the projectors between 'seg1' and 'seg2' with $P_{xy}^{(1)}$ and $P_{xy}^{(1b)}$ and between 'seg2' and 'seg3' $P_{xy}^{(2)}$ and $P_{xy}^{(2b)}$ (see Figure 1B). Note that we have to use two sub-ensembles: one with the set $P_{xy}^{(1)}$ and $P_{xy}^{(2)}$ and the other one with the set $P_{xy}^{(1b)}$ and $P_{xy}^{(2b)}$ for a successful compilation of the Coop-D-Ramsey sequence (see discussion in the SI and Figure S1). In addition, we design the system to end up in $|0\rangle$, not in $|+1\rangle$ as this is the case for the D-Ramsey.

We compared the performance of a D-Ramsey versus the Coop-D-Ramsey scheme in bulk diamond. To this end, we chose a single NV in bulk diamond and aligned an external weak magnetic field roughly along the NV axis. The splitting of the ESR transitions is 25.7 MHz. The Rabi frequencies in the reference configuration (relative amplitude $\kappa = 1$) of the energetically lower and higher spin transitions was 3.8 MHz and 3.5 MHz, respectively. We explain the slight difference by the frequency dependence of the microwave power of our apparatus. To determine the maximum spin contrast we performed a spin contrast measurement ($T_1$-Decay, see Figure 2D (black) ) as described elsewhere (see also description in Materials and Methods).[22] Then we acquired a Hahn-Echo also to verify that the coherence time of the NV center is long enough (Figure 2D (gray)). In the next step we measured the Coop-D-Ramsey and D-Ramsey by varying amplitude and detuning to simulate tumbling and temperature changes. To calculate the sequence we use the spin system as defined in eqs (S2a,b,c,d). Amplitude variations are accomplished by varying the relative amplitude $\kappa$ and detunings by varying $\Delta D$ (see also Figure 2C). The parameters for pulse compilation are given in the Materials and Methods part. We fitted every data set to a harmonic function with an exponential damping term and extracted the peak to peak amplitude (signal contrast). The signal shows an almost constant contrast within the amplitude and detuning variation limit that has been set for compilation ($\kappa$: 0.7 to 1.3, $\Delta D$: 0 to 200kHz). Interestingly, the sequence shows a better performance than the target limit. As one can see in Figure 2, the signal contrast of the optimal control sequence is almost twice as large as in the non-optimized case, whereas in case of detuning, the signal contrast is always better within a bandwidth of 1 MHz.

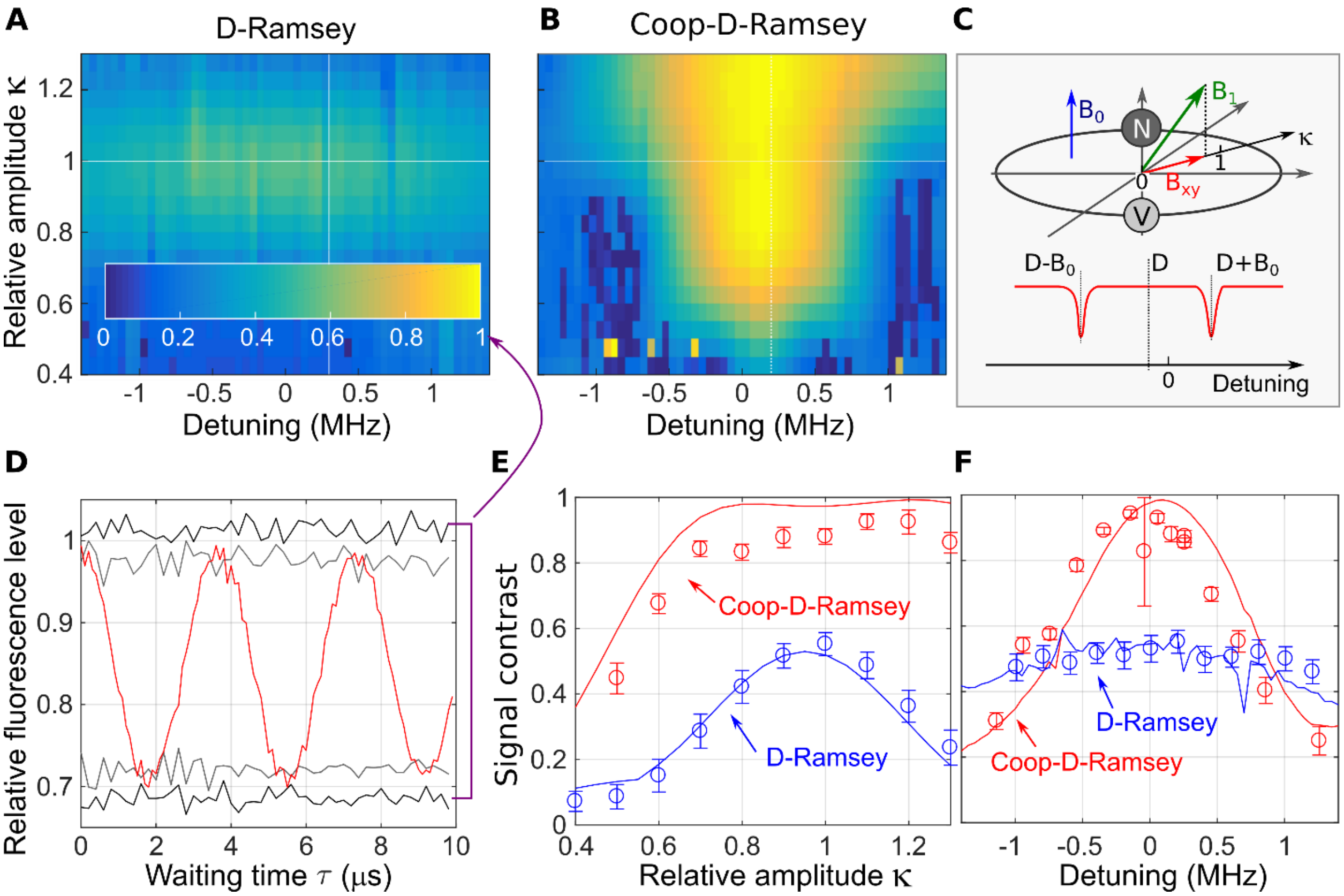

**Figure 2. Coop-D-Ramsey applied to a NV in bulk diamond. (A):** Simulation of D-Ramsey using square pulses. **(B):** Simulation of Coop-D-Ramsey for the settings as in (A). **(C):** Illustration of relative Amplitude $\kappa$ and detuning. **(D):** $T_1$-Decay (solid black), Hahn-Echo (solid gray) and Coop-D-Ramsey (red) for a slight detuning of around ~300 kHz. The purple bar represents the maximum signal normalized to 1. **(E)** and **(F):** comparison of the signal contrast by D-Ramsey (blue) and Coop-D-Ramsey (red) as a function of the relative amplitude $\kappa$ and detuning $\Delta D$. The solid lines represent simulations corresponding to the parameter slices marked with the white lines in (A) and (B). Circles with error bars are measured data. Note: If there is no description given for a specific scale, it is inherited from the graph along horizontal or vertical lines from left to right or top to bottom, respectively.

For the ND case we performed a similar analysis (see Figure 3). First, we compensated the external magnetic field to satisfy the condition $E \gg \beta_{0z}$ as required to use eqs (S3a,b,c,d) for modeling the NV spin system within a ND. Fitting a continuously driven optical detected magnetic resonance (cwODMR) spectrum (Figure S3) reveals $E = 4.35 \pm 0.01$ MHz and $\beta_{0z} = 0.53 \pm 0.03$ MHz. The reference Rabi frequencies on the lower and upper transition are 2.5 MHz and 2.8 MHz, respectively. Again, pulse parameters can be found in the Materials and Methods part.

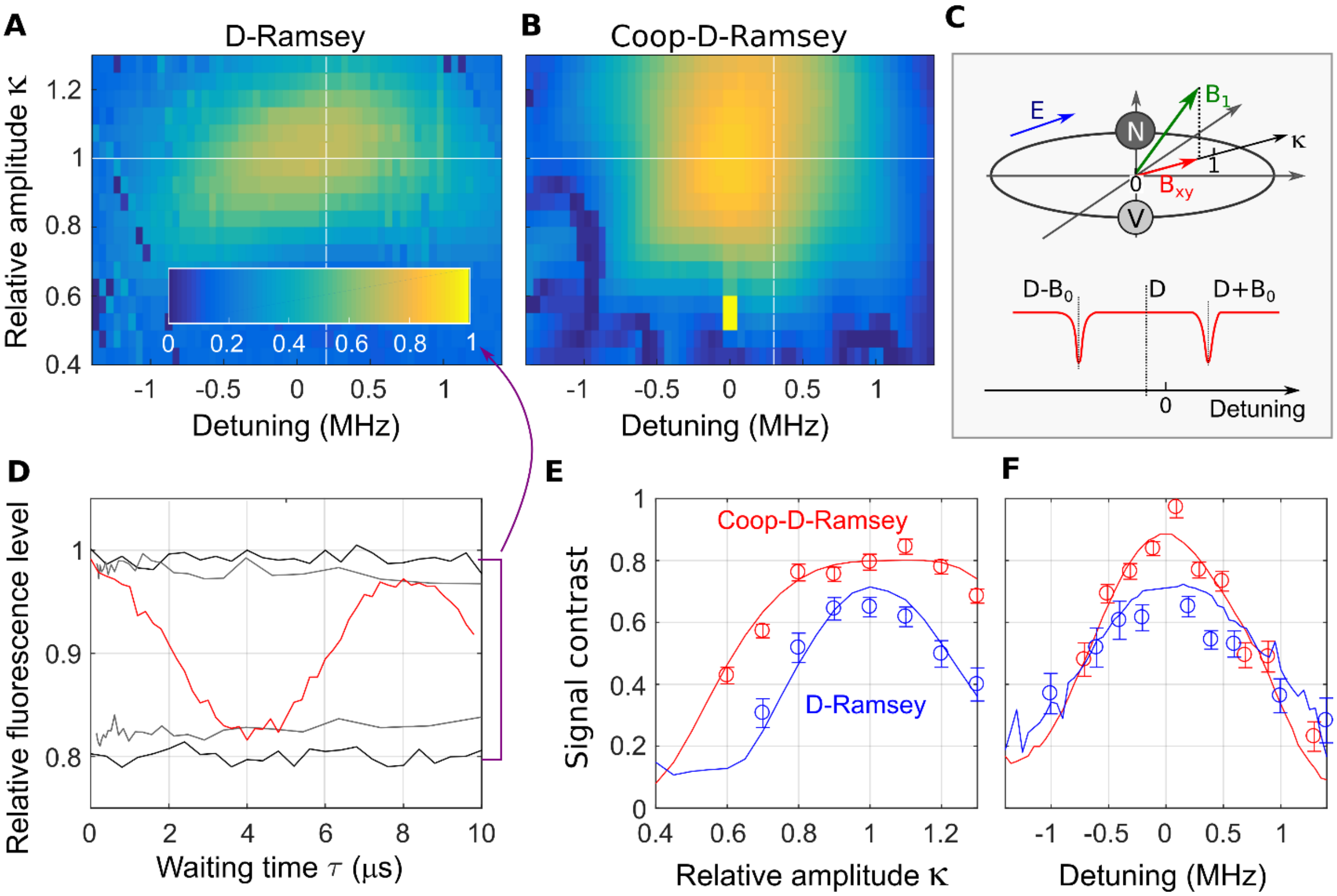

**Figure 3. Coop-D-Ramsey applied to a NV in a fixed ND. (A):** Simulation of D-Ramsey using square pulses. **(B):** Simulation of Coop-D-Ramsey for the settings as in (A). **(C)**: Illustration of the relative Amplitude $\kappa$ and detuning. **(D):** $T_1$-Decay (solid black), Hahn-Echo (solid gray) and Coop-D-Ramsey (red) for a slight detuning of ~125 kHz. The purple bar represents the maximum signal normalized to 1. **(E)** and **(F):** comparison of the signal contrast by D-Ramsey (blue) and Coop-D-Ramsey (red) as a function of the relative amplitude $\kappa$ and detuning. The solid lines represent simulations corresponding to the parameter slices marked with the white lines in (A) and (B). Circles with error bars are measured data. Note: If there is no description given for a specific scale, it is inherited from the graph along horizontal or vertical lines from left to right or top to bottom, respectively.

In contrast to bulk diamond the difference in signal contrast between D-Ramsey and Coop-D-Ramsey is less, but still significant. We attribute this difference to the reduction in the hyperfine splitting introduced by the zero-field parameter $E$: as the hyperfine lines are closer in frequency they are within the driving bandwidth of the conventional square pulses (see also Figure S2).

In a next step, we tested the cooperative pulse design on a tumbling nanodiamond. Therefore, we incorporated NDs in an agarose matrix, which acts as a cage for the NDs. The latter are spatially fixed but still can change their orientation over time. It turns out that NDs are typically immobilized in the beginning. We used this time window for characterizing the system to extract the Rabi driving strength, the zero field splitting $D$ and $E$ and the NV spin properties. After several hours of analysis, the NDs started to tumble on various timescales with rotational roundtrip time of several hours to milliseconds. Previously McGuiness *et al.*[12] observed tumbling of nanodiamonds in cells on similar timescales. We restricted our observations on NDs with slow angular changes compared to the overall sequence length, also because we want the driving strength to be constant during the application of one sequence interval (*i.e.* in agreement with the quasistatic parameter variation assumed for pulse engineering). In addition, we minimized the residual external magnetic fields by minimizing the linewidth and the COG of a cwODMR spectrum of a highly NV doped ND crystal (see Methods part in reference 1), as the COG position depends on the angle an external magnetic field is applied versus the NV axis. This would be interpreted as a temperature shift in case of a slowly tumbling ND (see Figure S4).

After compensating the external magnetic field, we tested the performance of the Coop-D-Ramsey on a ND doped with a single NV containing a $^{15}$N atom. We estimated $E$ to be around $0.2 \pm 0.1$ MHz and $|\beta_{0z}| = 0.2 \pm 0.1$ MHz in the beginning. The maximum Rabi frequency on the lower and upper transitions are $7.7 \pm 0.3$ MHz and $9.0 \pm 0.3$ MHz. Different to the former cases we optimized the pulse for every combination of $\kappa$ for the two transitions between 0.6 and 1.4 (see eq (S3d)).

After compiling a suitable sequence with DYNAMO, we continuously monitored the shift of the zero field splitting via a Coop-D-Ramsey for almost one day. In addition, we interleaved measurements of Rabi oscillations on both electron transitions and the overall temperature of the sample holder via a thermistor (see Figure 4). The contrast of the Coop-D-Ramsey is around 80% in the beginning. Over almost two hours the zero field splitting $D$ shifted by around 100 kHz. This corresponds to a heating of the confocal cell dish of roughly 1.5 K. This heating was not observed by the thermistor as it is monitoring the environmental temperature only at the sample chamber.

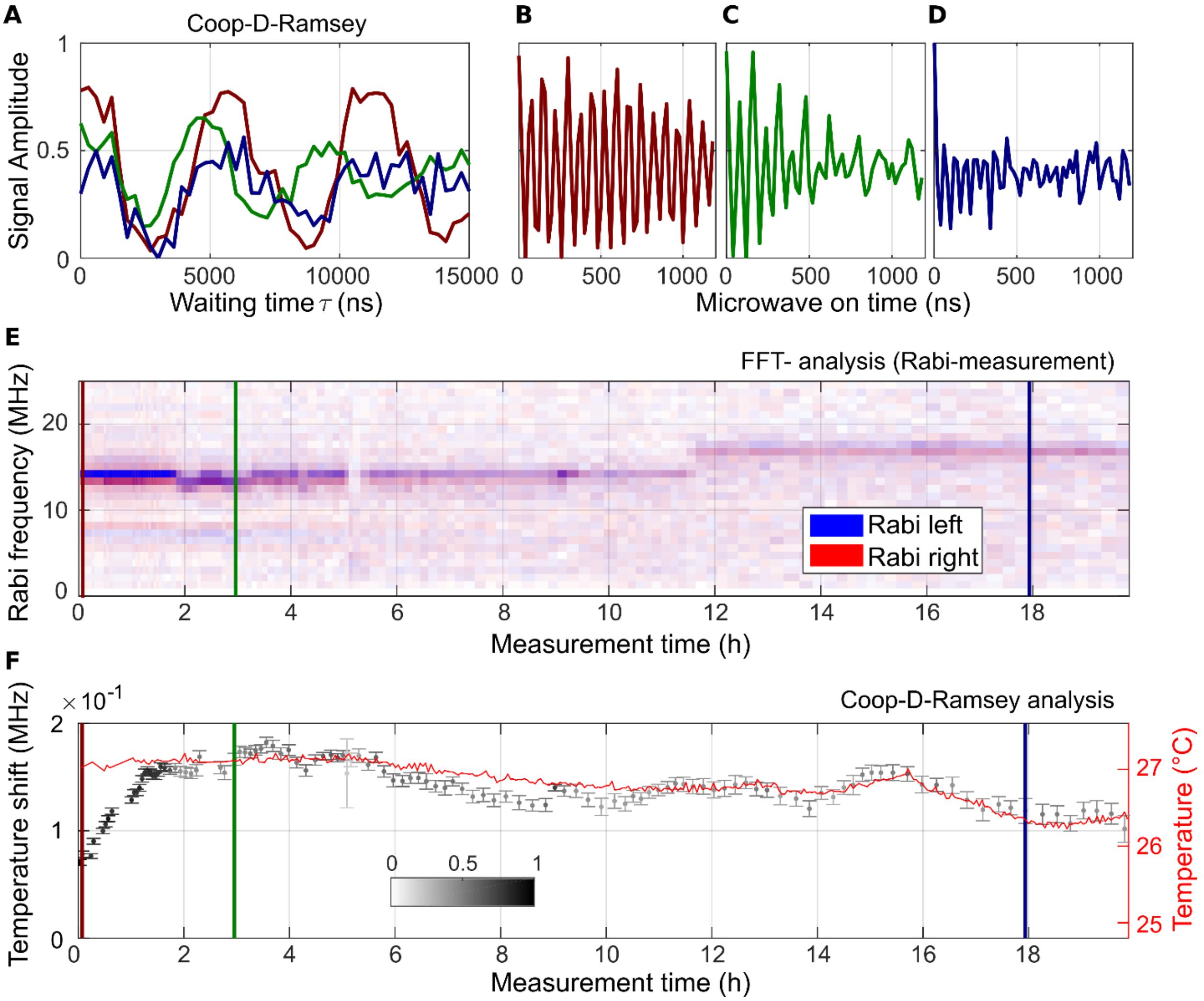

**Figure 4. Analysis of Coop-D-Ramsey for a rotating ND. (A):** Color coded Coop-D-Ramsey for single rotating NV at two different positions in time. The positions are marked in (E) and (F). **(B), (C)** and **(D):** Rabi oscillations driven on left transition. **(E):** FFT of Rabi measurement on left (blue) and right (red) electronic transition. For calculation of individual FFT data sets, single Rabi data have been bias corrected by its median amplitude. **(F):** Change of the zero field splitting measured via a Coop-D-Ramsey (white to black). To mark the signal contrast the individual data points are colored from white (zero contrast) to black (full contrast). In red the corresponding temperature read out of a thermistor is plotted. Note: that (B), (C) and (D) inherit their y-scale from (A).

During this heat-up phase the signal contrast of the Coop-D-Ramsey stayed almost constant, around 80%. After two hours the contrast dropped. Around 45 min later we reduced the detuning to $D$ by 100 kHz, to stay closer to the defined compilation interval (allowed detuning: 0-200kHz), also to rule out that the latter is the origin of the signal drop. A closer look on the acquired Rabi data in Figure 4E reveals that the FFT shows a significantly larger linewidth resulting from a fast decay of the Rabi oscillations. We interpret this behavior as an averaging of the NV having different orientations to the driving field during tumbling[23] (compare Figure 4B with C and D ). Furthermore, we acquired the autocorrelation function $g^{(2)}(\tau)$ before and after the end of the temperature measurement series (see Figure S5). Interestingly, the $g^{(2)}(\tau)$ measured after the temperature measurements does not approach one, *i.e.* the value for a totally uncorrelated signal. To find the timescale of the new occurring feature, we performed fluorescence correlation spectroscopy from sub-μs to sub-second timescale. The acquired $G(\tau)$ function reveals a bunching feature with a time constant of $\tau_l = 4.0 \pm 0.2$ ms. As we do not see a significant increase in width of the point spread function (Figure S5) in comparison to the measured point spread function for other fixed NDs, we identify the bunching feature in $G(\tau)$ and the offset in $g^{(2)}(\tau)$ to be a consequence of ND rotation.

Finally we compare the results for the Coop-D-Ramsey to the original D-Ramsey in case of rotating NDs. The D-Ramsey sequence requires by design the NV electronic transitions to be selectively addressable. This condition is not fulfilled by NDs, as their transitions are typical split in the range of ~kHz to ~MHz. [12,24,25] Nevertheless, we can estimate the maximum signal contrast expected for a rotating ND with the given parameters. To this end we simulated the D-Ramsey for various angles of the microwave field versus the NV axis and averaged the signal. Figure S6A in the SI shows the result of this averaging in case of a more bulk or ND like behavior. For the given parameter set (spin transition energy, Rabi amplitude and hyperfine splitting) we would not have been able to observe characteristic modulations corresponding to the shift in $D$ *i.e.* measure temperature.

There are alternative ways to measure temperature with rotating NDs. The simplest measurement scheme is using continuous wave, which, however, is limited by $T_2^*$.[26] In addition, the time resolution of the method is estimated to be on the order of several tens of ms.[12] In case of the D-Ramsey scheme one could also replace conventional pulses, by adiabatic passage.[27] For example, chirped pulses, allowing to create a state inversion with high fidelity and bandwidth, or the $B_1$ insensitive rotation (BIR) pulse, which allows to adjust a superposition state with arbitrary state composition. Here $B_1$ is the driving microwave field. Even if adiabatic pulses provide a robust way to manipulate the state of a spin system, the pulses itself have to be much longer than the corresponding effective Larmor frequency in the rotating reference frame to fulfill the adiabatic condition. This has to be the case even for the smallest $B_1$ field amplitude considered. In addition, the BIR pulse scheme can only provide a certain bandwidth in frequency space roughly in the range of the applied Rabi frequency.[27] Especially in bulk diamond and for a NV with $^{14}N$, the pulse has to have a bandwidth of around 4 MHz at least to drive all hyperfine lines, setting the lower limit of the Rabi driving strength. In the case of NDs, strain usually reduces the hyperfine splitting that shall lead to an increase in the overall performance of the BIR pulses. The problem of overall pulse length still remains.

The used cooperative design gives some advantages over conventional optimal control. In the latter case, each pulse of a sequence would be synthesized as a particular gate, which requires more resources (*e.g.* time or energy) as synthesis of a particular state transfer. In our cooperative design, not even a particular state transfer is required but just one out of a subset (e.g. any equal superposition state). Then, all different pulses in our sequence can correct errors in state adjustment of other pulses. A particular final state is achieved by cooperative adjustment of all pulses. This introduces an additional degree of freedom, allowing the algorithm to be more efficient concerning the overall length of the sequence,[19] providing better results for shorter pulses.

In this work we have chosen an overall pulse length of around 2 μs in case of the rotating ND. In case of a simulated echo, which essentially consist of two identical sequences split by a correlation interval, the shortest echo sequence would be 4 μs long. The maximum bandwidth one can access in terms of temporal resolution to measure temperature fluctuations, is thus around 250 kHz. This is comparable to the time resolution reported previously[28] where the resonance frequency at ~840 kHz of a mechanical oscillator was probed. Therefore, it should be possible to resolve temperature fluctuations with μs-time and potentially even nm-spatial resolution.

This might be of particular importance as recent experiments on living cell using fluorescent probes [29–33] have measured sizeable temperature gradients. Thermodynamic arguments however suggest, [34] that average temperature changes by endogenous thermogenesis should we very small, but do not exclude, that measurable temperature fluctuations may still exist. For example, Inomata *et al.*[28] observed a characteristic change of heat generation from a burst-like to a continuous increasing behavior after stimulating brown fat cells with norepinephrine in a well isolated chamber design.

In addition, NDs may be placed on a small structure whose size is sub-micron down to several tens of nanometers, to analyze their thermal dynamics. In difference to scanning probe approaches, where a sharp but microscopic tip is used to probe the local heat dissipation,[35,36] NDs can reach dimensions even below 10 nm and still host NV.[24,25] As diamond itself has a high thermal conductivity and a low heat capacity, high thermal coupling and low perturbation of the device under study is possible in direct contact. Hence high spatial resolution can be combined with fast temporal response, enable by the present measurement scheme.

## CONCLUSION

In summary, we have converted the D-Ramsey scheme for sensitive temperature measurements into a more robust Coop-D-Ramsey, which is based on optimal control theory in combination with a cooperative pulse design. We demonstrate superior robustness against variation in the driving strength and resonance mismatching, compared to conventional soft or hard pulses. The recipe we exploited to enable robust spin control is universal and can be used to design other pulse schemes.

In addition, we show that the Coop-D-Ramsey even performs well, when a nanodiamond is tumbling within an agarose matrix. Ultimately our work paves the way to measure sub 100 mK temperature fluctuations on microsecond time and nanometer length scale.

ACKNOWLEDGMENT

We acknowledge the support of the European Commission Marie Curie ETN "QuSCo" (GA N°765267), the KIST Open Research Program (2E27801) and the German science foundation (SPP 1601). In addition, this work was supported by ERC grant SMel, the Max Planck Society and the Humboldt Foundation and the EU - FET Flagship on Quantum Technologies through the Project ASTERIQS. We thank Thomas Schulte-Herbrüggen and Steffen Glaser for fruitful discussions.

## MATERIALS AND METHODS

**Sample preparation.** Bulk study: We glue a bulk diamond containing single deep NV on a printed circuit board as sample holder. The latter is also used to apply microwave excitation via a 50 µm thin copper wire spanned over the diamond.

ND study: For fixed nanodiamonds we spin coat a ND sample ("M19-S11c", SN: AA00M7, Diamond Nanotechnologies Inc.) onto a plasma cleaned cover glass glued on a PCB board. A 50 µm thin copper wire is spanned over the glass and connected to the PCB board for microwave excitation.

For tumbling nanodiamonds we modified the above approach by adding a sample chamber on top. We glued a confocal cell culture dish without the glass slide at the bottom onto the cover glass after spanning the wire. The sample is a 25 µl of ND stock solution ("M19-S11c", SN: AA00M7, Diamond Nanotechnologies Inc.) mixed with 25 mg agarose (target concentration 5 wt%, Agarose Typ 1-A A0169, Sigma) in 475 µl water. After dissolving the agarose under vigorous mixing and heating, the sample is applied to the dish. To prevent the agarose matrix from drying out 3 ml of water is added.

**Pulse parameters**. In the bulk sample case we compile the Coop-D-Ramsey sequence with a variation in possible detunings $\Delta D$ from zero to 200 kHz in 100kHz steps. The relative driving amplitude $\kappa$ is varied simultaneously from 0.7 to 1.3 in steps of 0.1. The overall sequence is 2640 ns long. We also add the hyperfine lines by including a small magnetic detuning as an additional ensemble. For the fixed nanodiamonds we choose the same D-shifts. The relative Rabi driving amplitude $\kappa$ is varied from 0.8 to 1.2 in steps of 0.1. The overall pulse length is 3840 ns. For rotating nanodiamonds we again choose the very same shifts in D. Different to former cases the amplitude is varied for every combination of left and right driving strength between 0.6 and 1.4 from its original values. The overall pulse length is 2160 ns. In both ND studies we include a small magnetic detuning, whereas we treat the central (fixed ND) or the center of all hyperfine transitions (rotating ND) for one electronic transitions as $E$ in eq (S3b).

**Experimental setup.** To test the Coop-D-Ramsey we use a home built confocal setup with a 532 nm laser to excite NV. To separate the strong laser light from NV fluorescence, we use a long pass filter with cutoff design frequency at 647 nm. For NV fluorescence detection we used two avalanche photo diodes (APDs) in a Hanbury-Brown-and-Twist configuration. For manipulation of the NV spin system two arbitrary waveform generators (AWG2041, Tektronics) working in Master/Slave mode are connected to the I and Q channel of an IQ-mixer (IMOH-01- 58, Pulsar Microwave). The local oscillator input is connected to a frequency source (SMIQ03B, Rohde&Schwarz). The IQ mixer is connected to a microwave switch (ZASWA-2-50DR+, mini-circuits) allowing to additionally suppress microwave excitation. After the switch the microwave excitation is amplified by a 16W amplifier (ZHL-16W- 43+, mini-circuits), which is directly connected to the sample. For microwave excitation a simple copper wire is used. A thermistor was connected to the sample holder, to monitor the temperature of the sample. The setup is fully controlled via custom made software programs written in Python. To record the fluorescent autocorrelation function $G(\tau)$, one of the APDs is connected to a time-correlated single photon counting module (PicoHarp 300, PicoQuant).

**Data analysis**. Data analysis, simulations and data visualization have been performed in Matlab (Mathworks). To compare the measured data with corresponding simulations we have to extract the individual spin contrast of the NV under study. To this end the spin of NV is first initialized and fluorescence is readout after a sufficient long time interval with green laser pulses. The fluorescence response of NV is depending on its spin state: The first ~300 ns after turning on the laser give a higher fluorescence signal if being in $|0\rangle$ and lower if being in $|\pm 1\rangle$. We perform two spin contrast measurement: the first one is done without any microwave excitation between initialization and readout, giving the reference value for being in $|0\rangle$. In a second run we depopulate $|0\rangle$, using a linear chirp microwave pulse and extract the reference value for being in $|\pm 1\rangle$. In the bulk and fixed ND case we determine the spin contrast in a separate measurement. For the rotating nanodiamond study the spin contrast is continuously monitored during the running temperature readout. Further we use DYNAMO to simulate the state evolution of individual NV when varying the free evolution time $\tau$ for a (Coop-)D-Ramsey. As we do not include decoherence effects at this stage, and as we want to compare simulated and experimental results, we also measure a Hahn Echo for the individual NVs. The Hahn Echo is then fit to a stretched exponential function, which parameters are used as an exponential damping term to fit experimental data. In case of the rotating ND we keep the parameters of the stretched exponential as a free fit parameter.

To extract the correlation time of rotating nanodiamond we use a mono exponential decay. Figure illustrations have been accomplished using Inkscape.

## SUPPORTING INFORMATION

**Description of spin system for pulse compilation.** To compile a Coop-D-Ramsey sequence, one needs to describe the NV quantum system in terms of the system Hamiltonian and the set of parameters, which are varied by the tumbling motion of the nanodiamonds. We define $H_{\text{drift}}$ and $H_{\text{c}} = \sum_k u_k(t)\, H_k$ (see main text) as:

$$H_{\text{drift}} = D \cdot S_z^2 + E\left(S_x^2 - S_y^2\right) + \sum_{k=x,y,z} \beta_{0k} S_k \tag{S1a}$$

and

$$H_{\text{c}}(t) = \sum_{k=x,y,z} \beta_{1k}(t) S_k. \tag{S1b}$$

$D$ is the zero field splitting, $E$ is the zero field component introduced by off-axial strain, $S_k$ the spin operators of a spin one system, $\beta_{0k}$ static components introduced by an external magnetic field with strength $\beta_0 = \sqrt{\sum_k \beta_{0k}^2}$ , and $\beta_{1k}$ the components of the microwave field.

We distinguish two cases. The first case is a NV with $|E| \ll \beta_0$, when the external static field shall be aligned along the $z$-axis: $|\beta_{0z}| = \beta_0$. We call this bulk case. And second, an NV with $|E| \gg \beta_0$, assigned as the ND case. The microwave field is chosen to be an in-plane linear polarized field with angle $\phi$ to the $x$-axis ($\beta_{1x}(t) = \beta_1 \cos(\omega t + \phi)$, $\beta_{1y}(t) = \beta_1 \sin(\omega t + \phi)$ and $\beta_{1z}(t) = 0$ ). Transforming eqs (S1a) and (S1b), into a left-handed and a right-handed rotating reference frame, leads to the overall system Hamiltonian $H$ for the bulk case using the assumption $D \gg \beta_1$:

$$H = H_{\text{drift}} + u_x(t) \cdot X + u_y(t) \cdot Y, \tag{S2a}$$

with

$$H_{\text{drift}} = \begin{pmatrix} \Delta D + \beta_{0z} & 0 & 0 \\ 0 & 0 & 0 \\ 0 & 0 & \Delta D - \beta_{0z} \end{pmatrix} \tag{S2b}$$

and

$$X = \frac{\beta_1}{2} \begin{pmatrix} 0 & 1 & 0 \\ 1 & 0 & 1 \\ 0 & 1 & 0 \end{pmatrix}, Y = \frac{\beta_1}{2} \begin{pmatrix} 0 & -\mathrm{i} & 0 \\ \mathrm{i} & 0 & \mathrm{i} \\ 0 & -\mathrm{i} & 0 \end{pmatrix}. \tag{S2c}$$

Note that the described relations in eqs (S2) are valid even for a residual microwave field aligned along the z-axis. As $D$ is typically 2.87 GHz for NV and the Rabi amplitude is on the order of several MHz the fast fluctuations the $S_z$ operator introduces are averaged out.

To allow a variation in the driving strength of the microwave field, we introduce the relative amplitude $\kappa$.The controls $X_l$ and $Y_l$ for a single set $l$ of the whole parameter set to optimize for, are modified to:

$$X_l = \kappa(l) \frac{\beta_1}{2} \begin{pmatrix} 0 & 1 & 0 \\ 1 & 0 & 1 \\ 0 & 1 & 0 \end{pmatrix} \text{ and } Y_l = \kappa(l) \frac{\beta_1}{2} \begin{pmatrix} 0 & -\mathrm{i} & 0 \\ \mathrm{i} & 0 & \mathrm{i} \\ 0 & -\mathrm{i} & 0 \end{pmatrix}. \tag{S2d}$$

In real experimental situation $\beta_1$will be the measured Rabi frequency that is determined before pulse compilation.

In the ND case the situation becomes more complicated:

$$H = H_{\text{drift}} + u_x(t) \cdot [\cos(\phi)\, X_{\parallel} + \sin(\phi) X_{\perp}] + u_y(t) \cdot [\cos(\phi)\, Y_{\parallel} + \sin(\phi) Y_{\perp}] \tag{S3a}$$

with:

$$H_{\text{drift}} = \begin{pmatrix} \Delta D + E & 0 & 0 \\ 0 & 0 & 0 \\ 0 & 0 & \Delta D - E \end{pmatrix}, \tag{S3b}$$

and:

$$X_{\parallel} = \frac{\beta_1}{\sqrt{2}} \begin{pmatrix} 0 & 1 & 0 \\ 1 & 0 & 0 \\ 0 & 0 & 0 \end{pmatrix}, Y_{\parallel} = \frac{\beta_1}{\sqrt{2}} \begin{pmatrix} 0 & -\mathrm{i} & 0 \\ \mathrm{i} & 0 & 0 \\ 0 & 0 & 0 \end{pmatrix}, X_{\perp} = \frac{\beta_1}{\sqrt{2}} \begin{pmatrix} 0 & 0 & 0 \\ 0 & 0 & 1 \\ 0 & 1 & 0 \end{pmatrix}, Y_{\perp} = \frac{\beta_1}{\sqrt{2}} \begin{pmatrix} 0 & 0 & 0 \\ 0 & 0 & \mathrm{i} \\ 0 & -\mathrm{i} & 0 \end{pmatrix}. \tag{S3c}$$

Again, for parameter variation eq (S3c) is redefined as:

$$X_l = \frac{1}{2} \begin{pmatrix} 0 & \beta_{1_{\parallel}} \kappa(l) & 0 \\ \beta_{1_{\parallel}} \kappa(l) & 0 & \beta_{1_{\perp}} \kappa'(l) \\ 0 & \beta_{1_{\perp}} \kappa'(l) & 0 \end{pmatrix}, Y_l = \frac{\mathrm{i}}{2} \begin{pmatrix} 0 & -\beta_{1_{\parallel}} \kappa(l) & 0 \\ \beta_{1_{\parallel}} \kappa(l) & 0 & \beta_{1_{\perp}} \kappa'(l) \\ 0 & -\beta_{1_{\perp}} \kappa'(l) & 0 \end{pmatrix}. \tag{S3d}$$

Now $\beta_{1_{\parallel}}$ and $\beta_{1_{\perp}}$ are directly the measured Rabi frequency of left and right ESR transitions. In case of the fixed ND study $\kappa(l)$ equals $\kappa'(l)$. For the rotating ND $\kappa(l)$ and $\kappa'(l)$ are varied individually, but $\beta_{1_{\parallel}}$ is set equal to $\beta_{1_{\perp}}$.

The interpretation of eq (S3c) can also be understood in the following way: If a strain field is aligned along the microwave field only one transition can be driven. If it is perpendicular aligned, one can only drive the other transition.

**Projectors to for cooperative pulse design.** To introduce cooperativity, we use the projector sets $P_{xy}^{(1)}$ / $P_{xy}^{(2)}$ (set1) and $P_{xy}^{(1b)}$/ $P_{xy}^{(2b)}$ (set2) as shown in eqs (S4a,b) in two sub-ensemble. All $P_{xy}$ project into a 2-dimensional subspace of the Liouville space of density matrices, with the difference, that Projectors with the super script (b) introduce an additional $\frac{\pi}{2}$ phase shift. The reason to do so, is that we want to account for the situation, where some phase is picked up during the free evolution intervals with length $\frac{\tau}{2}$. The minimum set to account for all phase accumulations between zero and 2π is covered by both. Note that $P_{xy}^{(1b)}$ adds a positive phase, whereas $P_{xy}^{(2b)}$ the corresponding negative one. If we would have chosen a positive phase in the latter the final state must be $|+1\rangle$ instead of $|0\rangle$.

$$P_{xy}^{(1)} = \begin{pmatrix} 0 & & & & & & & & \\ & 0 & & & & & & & \\ & & 0 & & & & & & \\ & & & 0 & & & & & \\ 0.5 & & & & 0.5 & & & & 0.5 \\ & & & & & 1 & & 0 & \\ & & & & & & 0 & & \\ & & & & & 0 & & 1 & \\ 0.5 & & & & 0.5 & & & & 0.5 \end{pmatrix}, \quad P_{xy}^{(2)} = \begin{pmatrix} 0.5 & & & & 0.5 & & & & 0.5 \\ & 1 & & 0 & & & & & \\ & & 0 & & & & & & \\ & 0 & & 1 & & & & & \\ 0.5 & & & & 0.5 & & & & 0.5 \\ & & & & & 0 & & & \\ & & & & & & 0 & & \\ & & & & & & & 0 & \\ & & & & & & & & 0 \end{pmatrix} \tag{S4a}$$

$$P_{xy}^{(1b)} = \begin{pmatrix} 0 & & & & & & & & \\ & 0 & & & & & & & \\ & & 0 & & & & & & \\ & & & 0 & & & & & \\ 0.5 & & & & 0.5 & & & & 0.5 \\ & & & & & \mathrm{i} & & 0 & \\ & & & & & & 0 & & \\ & & & & & 0 & & -\mathrm{i} & \\ 0.5 & & & & 0.5 & & & & 0.5 \end{pmatrix}, \quad P_{xy}^{(2b)} = \begin{pmatrix} 0.5 & & & & 0.5 & & & & 0.5 \\ & \mathrm{i} & & 0 & & & & & \\ & & 0 & & & & & & \\ & 0 & & -\mathrm{i} & & & & & \\ 0.5 & & & & 0.5 & & & & 0.5 \\ & & & & & 0 & & & \\ & & & & & & 0 & & \\ & & & & & & & 0 & \\ & & & & & & & & 0 \end{pmatrix} \tag{S4b}$$

To demonstrate that it is mandatory to use both projector sets, we compile one sequence with $P_{xy}^{(1)}$ and $P_{xy}^{(2)}$ only, and one including all projectors.

The system was defined as described in eq (S2a,b,c). The shift introduced by an external z-magnetic field has been set to 5 MHz, resulting in a splitting of 10 MHz for the ESR transitions. In addition, DYNAMO shall optimize the pulse response for three different configurations that resample the hyperfine structure of $^{14}N$. Therefore additional magnetic shifts with 0 MHz and ± 2.16 MHz are introduced. To see how phase accumulates during the evolution interval, the detuning $\Delta D$ is set to 200 kHz. Doing so one would expect a single harmonic modulation between state $|0\rangle$ and $|+1\rangle$ only with a period of 5 µs matching the adjusted detuning, when varying the time $\tau$ of the evolution intervals. Simulating the state dynamics of the model NV for different evolution intervals reveal, that in case of set1 fast oscillations can occur for different hyperfine lines (see Figure S1A). When using set1 and set2, one can observe a clear oscillation between state $|0\rangle$ and $|+1\rangle$ with a period of 200 kHz for all hyperfine lines (Figure S1B). The additional frequency components between 1 MHz and 4 MHz disappeared in the corresponding Fourier transformation (compare Figure S1C and D).

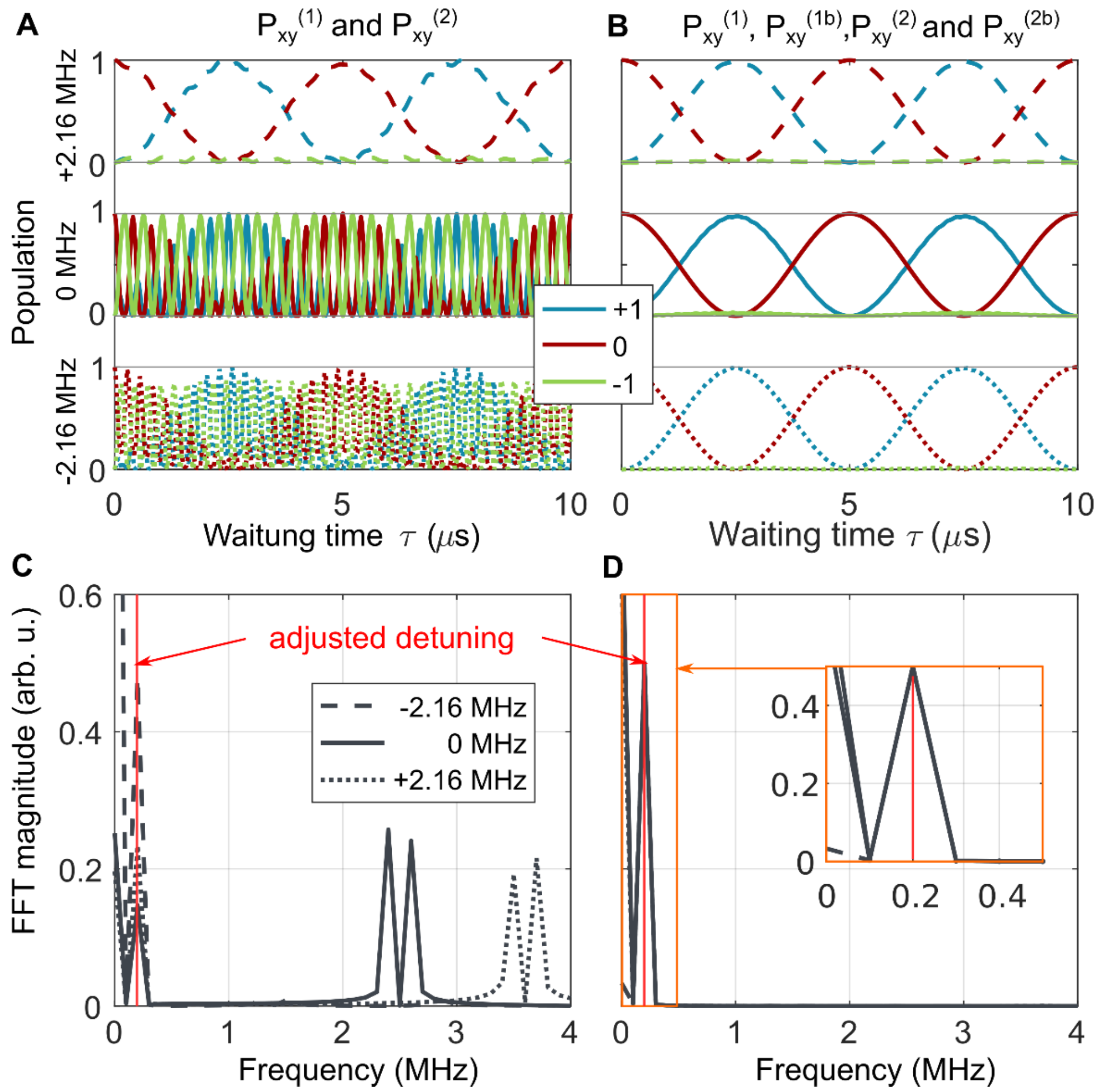

**Figure S1. Simulated evolution of Coop-D-Ramsey for different projector combinations: (A):** Evolution for NV spin states with and without $^{15}$N hyperfine when applying a Coop-D-Ramsey. Only Projector $P_{xy}^{(1)}$ and $P_{xy}^{(2)}$ are used to compile the pulse. **(B):** Compilation includes an additional sub-ensemble applying $P_{xy}^{(1b)}$ and $P_{xy}^{(2b)}$ instead of Projector $P_{xy}^{(1)}$ and $P_{xy}^{(2)}$. **(C):** Fast Fourier transformation (FFT) of the $|0\rangle$ population shown in (A) with and without hyperfine. **(D):** FFT of the $|0\rangle$ population shown in (B) with and without hyperfine. In (C) and (D) also the adjusted detuning $\Delta D$ is drawn by a red line.

**Hyperfine transitions of strained nanodiamonds.** To calculate the ESR transition of the nitrogen vacancy center in nanodiamonds we use following Hamiltonian:

$$H_{\mathrm{NV}} = D \cdot S_z^2 + E\left(S_x^2 - S_y^2\right) + \beta_0 \cdot S_z + S \cdot A \cdot I. \quad \text{(S5)}$$

Whereas $I$ are the nuclear spin matrices and $A$ the hyperfine interaction tensor[37]. For reasons of simplification we consider the $S_z I_z$ hyperfine coupling only.

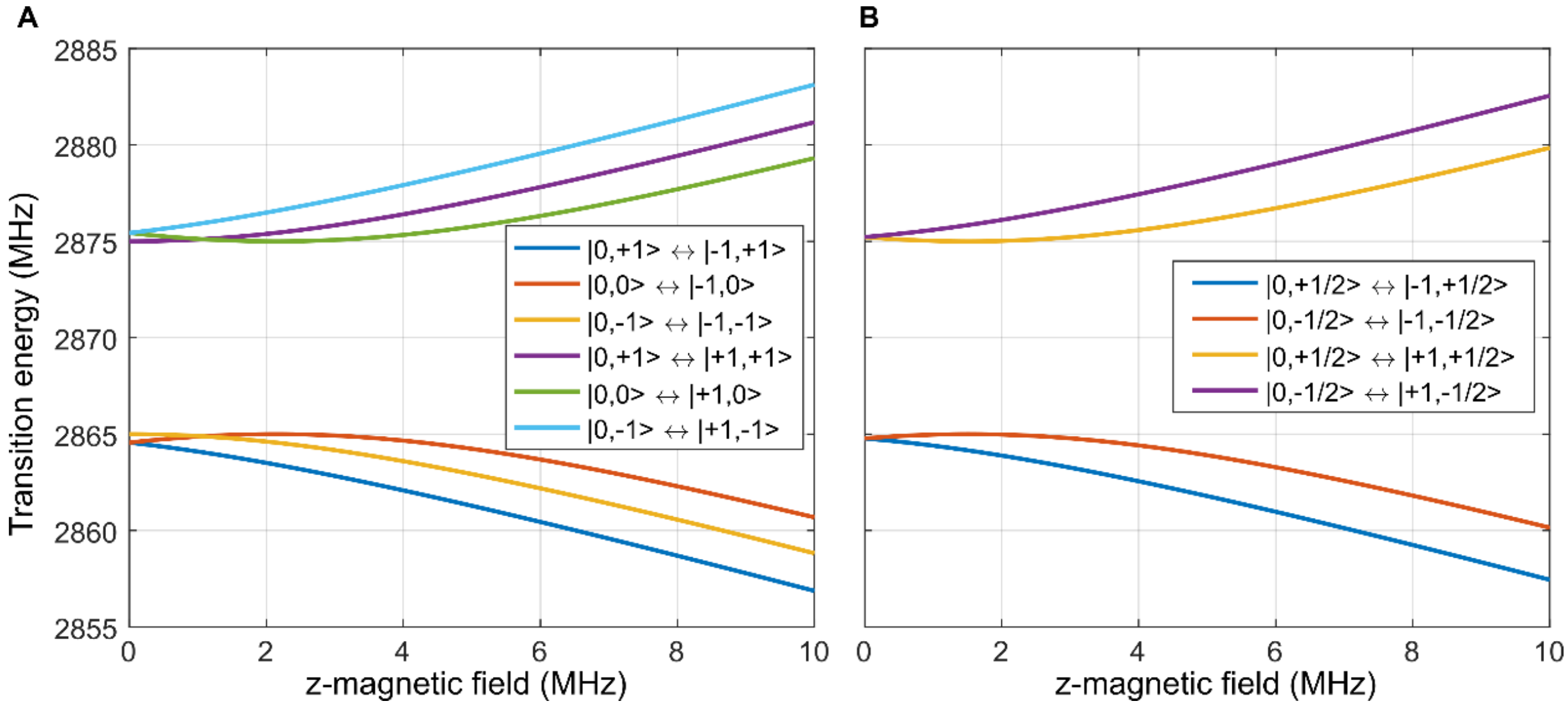

**Figure S2. Hyperfine transitions of a NV$^-$ in a ND in case of $^{14}$N (A) and $^{15}$N (B).** Note that the plotted transitions between states have different Eigen-basis in the region $E \approx \beta_0$, but for a better visibility the transitions are labeled for the limit $E \ll \beta_0$.

Then, we calculate the Eigensystem of $H_{\mathrm{NV}}$ and determine the most prominent transitions introduced by the spin operators $S_x$ and $S_y$ in the transformed system. Figure S2 shows this calculation for varying $\beta_0$ for a zero field splitting $E$=5 MHz and $^{14}$N. As one can see the hyperfine lines of the $m_I = \pm 1$ state overlap for $\beta_{0z} \ll E$. With increasing magnetic field the lines begin to separate. For $\beta_{0z} \geq \sim \frac{E}{2}$ the hyperfine lines are well separated. In Figure S2B we plot the corresponding transitions in case of $^{15}$N.

As an example Figure S3 shows the cwODMR spectrum of the ND used for the ND case study in the main text. The residual magnetic field along the z-axis correspond to $0.53 \pm 0.03$ MHz. The zero field parameter $E$ equals $4.35 \pm 0.01$ MHz.

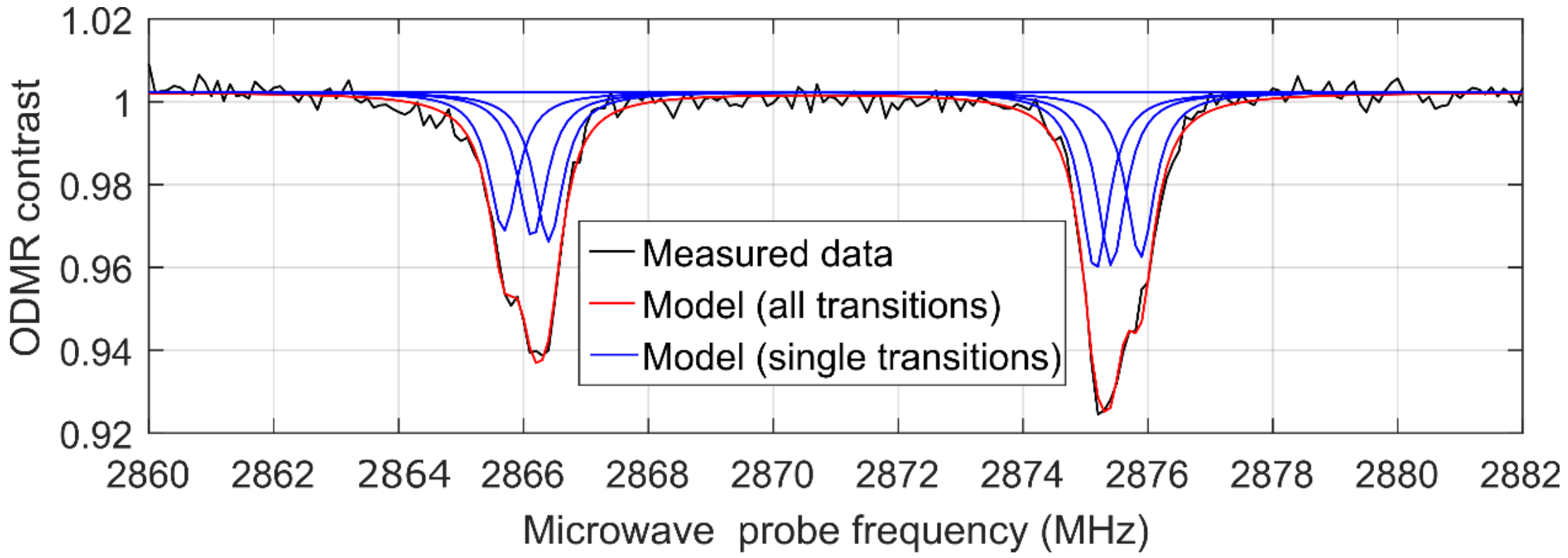

**Figure S3. cwODMR of a nanodiamond exhibiting strain.** The acquired data set (black) is fit to a model on basis of eq (S5). The red line shows the sum of all transitions whereas the blue ones show the individual transitions.

**Precision in temperature reading for a slow rotating ND.** The nitrogen vacancy can be used to measure temperature. The reason for the latter lies within the temperature dependence of the zero field splitting $D$ that can be approximated to be linear at ambient conditions. Thereby the shift in $D$ is around $70 - 80$ kHz $\mathrm{K}^{-1}$ [20]. For comparison the shift of a single spin transitions of NV is around 2.8 MHz $\mathrm{G}^{-1}$. So already for the earth magnetic field with ~0.5 G a line shift of ~1.4 MHz is expectable and would correspond to a temperature shift around 20 K. Small changes in the sub-G regime may already be interpret as a temperature change. Therefore the D-Ramsey is designed in a way that magnetic shifts will be partially compensated by measuring the center of gravity (COG) of both transitions of NV. If both transition move in the opposite direction in energy space (for example by a magnetic field aligned along the NV spin axis) the COG does not change. This is not always the case. Changing the orientation of an external magnetic field versus the NV axis lead to a modulation of the COG of both lines (see Figure S4A and S4B for an external field of 10 G). For very slow tumbling NDs its rotation may therefore be interpreted as a change in temperature. We define the maximum temperature uncertainty by the taking the maximum delta in the shift of the COG divided by 74.2 kHz $\mathrm{K}^{-1}$ and plot it in dependence of the external magnetic field (see Figure S4C). For an external field of 1 G an uncertainty in temperature reading of 55 mK is reached.

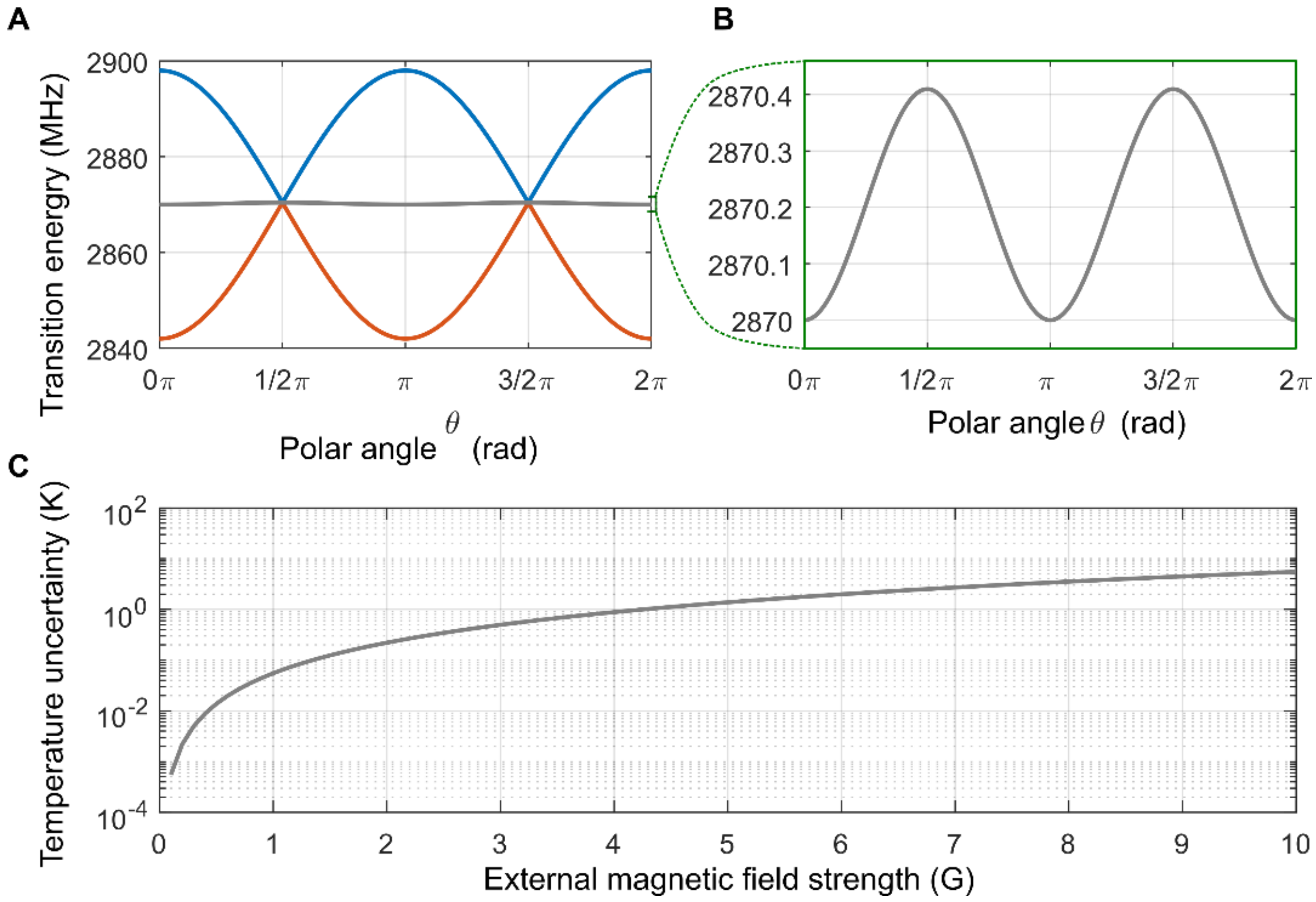

**Figure S4. Angular dependence of electronic transitions if an external magnetic field is applied. (A):** The polar angle $\theta$ of an external magnetic field of 10 G versus the NV axis is changed. (Blue) and (red) represent the upper and lower ESR transition, respectively. (Gray) present the center of gravity. **(B):** Zoom into (A) to see the center of gravity of both transitions. **(C):** For a fixed polar angle of $\frac{\pi}{2}$ the external magnetic field is ramped. The maximum possible difference of the COG, which is the difference between $\theta = \frac{\pi}{2}$ and $\theta = 0$, is plotted in terms of a temperature shift. For the latter we used $\frac{\partial D}{\partial T} = 74.2\ \mathrm{kHz\ K^{-1}}$ .

**Analysis of rotating ND.** As state in the main text we use an agarose matrix to restrict transversal diffusion. Figure S5 illustrates the behavior of the used NDs in the gel matrix. For a single NV one typically utilizes the correlation function $g^2(\tau)$ to reveal its single photon emission characteristics. The $g^{(2)}(\tau)$ function of the rotating ND in the main text is shown in Figure S5A (blue dots).

Interestingly we observe a characteristic change of $g^{(2)}(\tau)$ after several hours of temperature read out (Figure S5A black dots). A new bunching feature occurs, visible in a biasing of the correlation value for long $\tau$ to higher values. Analyzing the fluorescence signal using the so called $G(\tau)$ function, reveals this bunching to have a characteristic time scale of $\tau_l = 4.0 \pm 0.2$ ms.

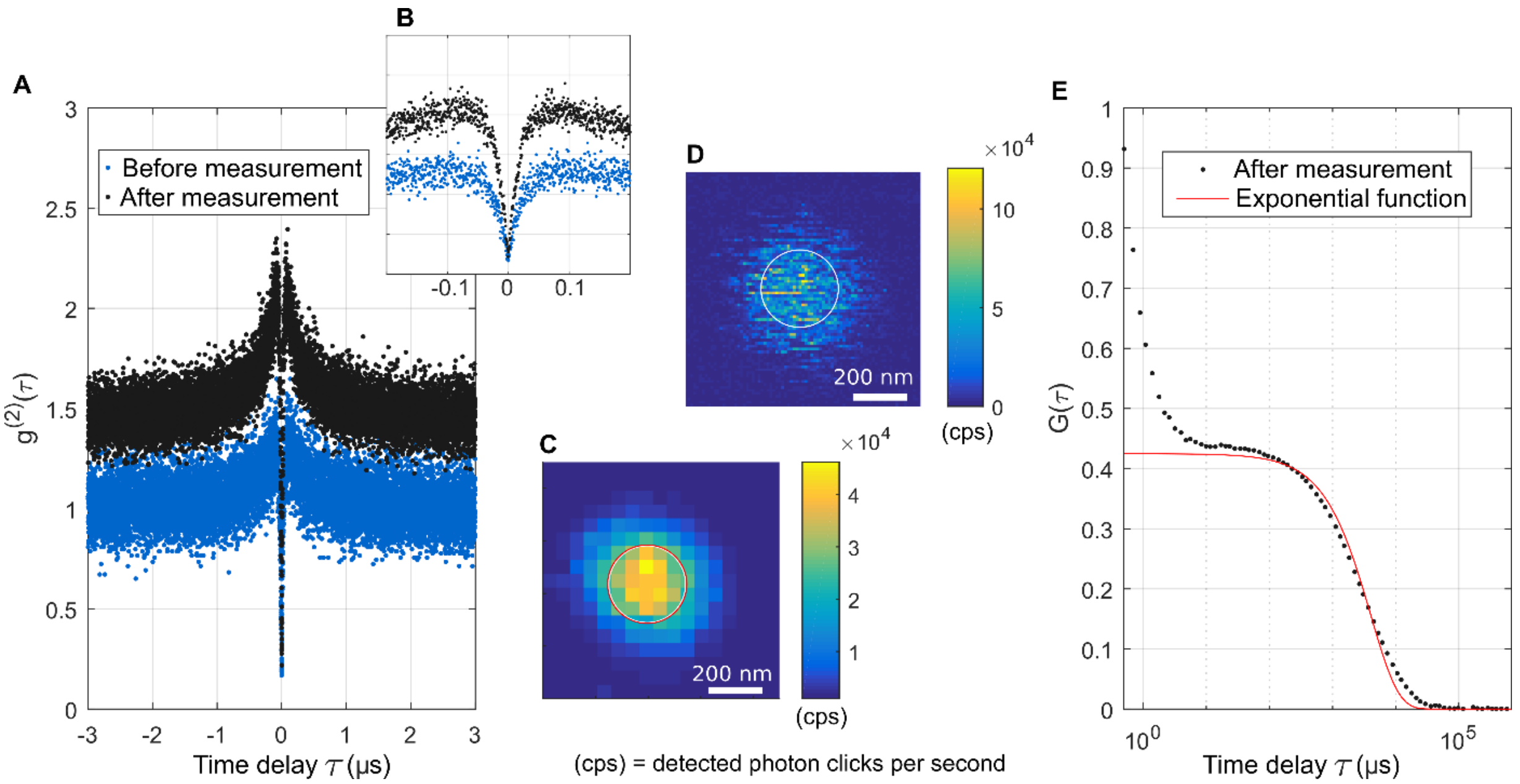

**Figure S5: Analysis of rotating ND: (A):** autocorrelation function $g^2(\tau)$ before (blue) and after (black) temperature measuring series. **(B):** Zoom in to the first 200 ns time delay of (A). **(C)** and **(D):** confocal scan after temperature measuring series with an integration time of 500 ms (C) and 5 ms (D). **(E):** Autocorrelation function $G(\tau)$ (black). The red line indicates a mono exponential fit to $G(\tau)$ for a time delay greater than 20 μs. The white cycles in (C) and (D) correspond to the measured average size of the point spread function for 10 different spots. The red cycle in (C) is the fitted average size of the shown confocal spot. The y-scaling of (B) is the same as (A), whereas the x-scale is the photon delay time in μs.

As the carrier of information in case of photon emission is the fluorescence intensity and its dynamics, we should see similar intensity fluctuation, if we perform a *xy*-scan with a suitable high temporal resolution. To this end, we perform a fast *xy*-scan (Figure S5D). If one acquires a confocal *xy*-map with sufficient slow scan speed, the size of the confocal spot still corresponds well to the measured average point spread function of other fluorescent spots (compare white and red cycle in Figure S5C). Therefore we interpret the occurring intensity fluctuations to be originated from the rotation of the ND, rather than transversal diffusion, which would also lead to fluctuations of the fluorescence signal.

**Simulating the D-Ramsey or a tumbling nanodiamond.**

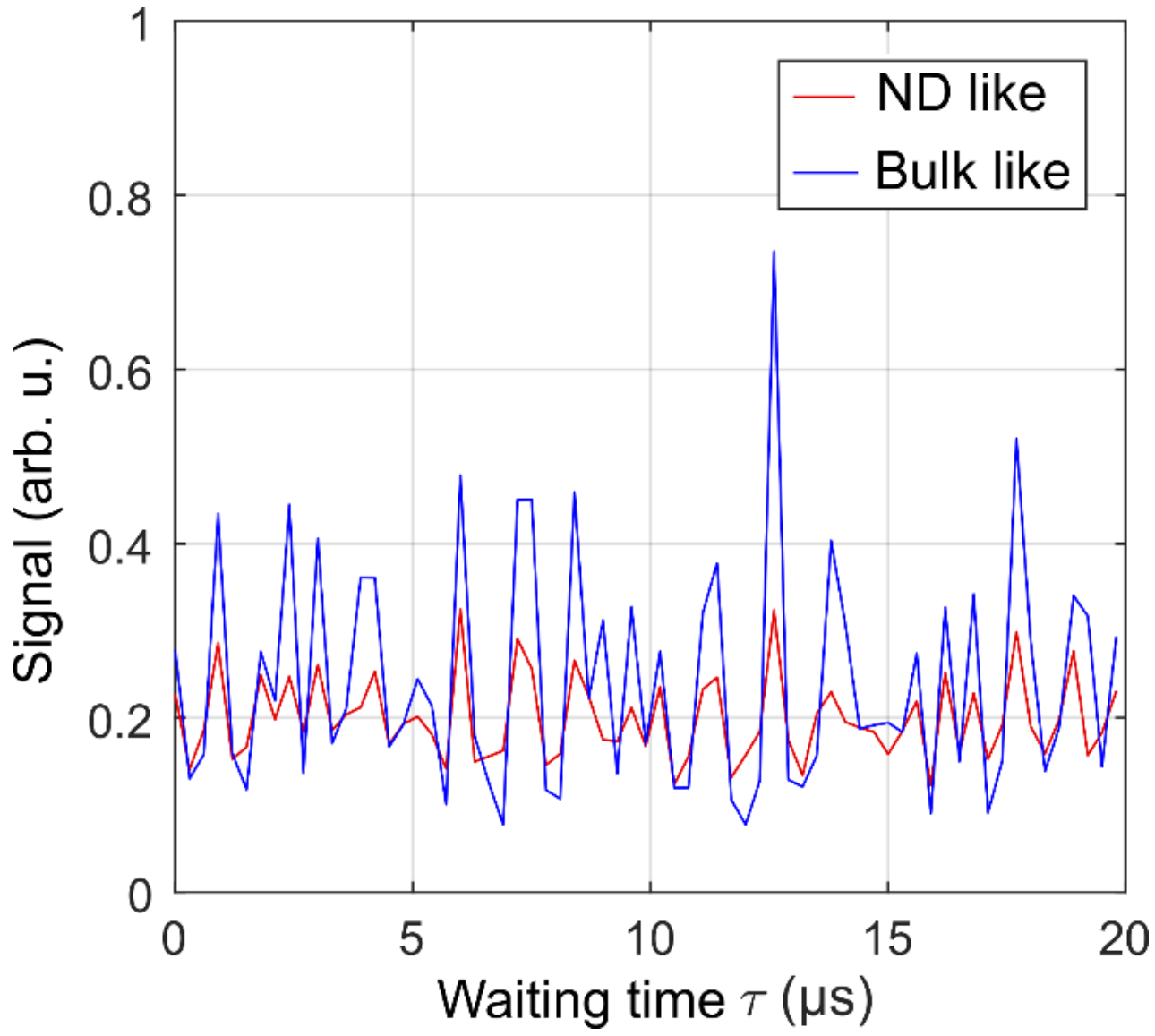

**Figure S6: Simulating the D-Ramsey or a tumbling nanodiamond:** For all simulations the ESR parameters for the rotating ND in the main text have been chosen. In ND case (ND like) eqs (S3) have been used. For the bulk case (Bulk like) eqs (S2). The simulated detuning was set to 200 kHz.